\documentclass[showpacs,preprintnumbers,amsmath,amssymb]{revtex4}
\usepackage{docs}

\usepackage{graphicx}
\usepackage{dcolumn}

\usepackage{bm}

\begin{document}

\title{Phantom energy traversable wormholes}%

\author{Francisco S. N. Lobo}
\email{flobo@cosmo.fis.fc.ul.pt}

\affiliation{Centro de Astronomia
e Astrof\'{\i}sica da Universidade de Lisboa, \\
Campo Grande, Ed. C8 1749-016 Lisboa, Portugal}

\begin{abstract}

It has been suggested that a possible candidate for the present
accelerated expansion of the Universe is ``phantom energy''. The
latter possesses an equation of state of the form $\omega\equiv
p/\rho<-1$, consequently violating the null energy condition. As
this is the fundamental ingredient to sustain traversable
wormholes, this cosmic fluid presents us with a natural scenario
for the existence of these exotic geometries. Note, however, that
the notion of phantom energy is that of a homogeneously
distributed fluid. Nevertheless, it can be extended to
inhomogeneous spherically symmetric spacetimes, and it is shown
that traversable wormholes may be supported by phantom energy. Due
to the fact of the accelerating Universe, macroscopic wormholes
could naturally be grown from the submicroscopic constructions
that originally pervaded the quantum foam. One could also imagine
an advanced civilization mining the cosmic fluid for phantom
energy necessary to construct and sustain a traversable wormhole.

In this context, we investigate the physical properties and
characteristics of traversable wormholes constructed using the
equation of state $p=\omega \rho$, with $\omega<-1$. We analyze
specific wormhole geometries, considering asymptotically flat
spacetimes and imposing an isotropic pressure. We also construct a
thin shell around the interior wormhole solution, by imposing the
phantom energy equation of state on the surface stresses. Using
the ``volume integral quantifier'' we verify that it is
theoretically possible to construct these geometries with
vanishing amounts of averaged null energy condition violating
phantom energy. Specific wormhole dimensions and the traversal
velocity and time are also deduced from the traversability
conditions for a particular wormhole geometry. These phantom
energy traversable wormholes have far-reaching physical and
cosmological implications. For instance, an advanced civilization
may use these geometries to induce closed timelike curves,
consequently violating causality.

\end{abstract}


\maketitle

\section{Introduction}

It is extraordinary that recent observations have confirmed that
the Universe is undergoing a phase of accelerated expansion.
Evidence of this cosmological expansion, coming from measurements
of supernovae of type Ia (SNe Ia)~\cite{Riess2,Perlmutter} and
independently from the cosmic microwave background
radiation~\cite{Bennet,Hinshaw}, shows that the Universe
additionally consists of some sort of negative pressure ``dark
energy''. The Wilkinson Microwave Anisotropy Probe (WMAP),
designed to measure the CMB anisotropy with great precision and
accuracy, has recently confirmed that the Universe is composed of
approximately $70$\% of dark energy~\cite{Bennet}. Several
candidates representing dark energy have been proposed in the
literature, namely, a positive cosmological constant, the
quintessence fields, generalizations of the Chaplygin gas and
so-called tachyon models. A simple way to parameterize the dark
energy is by an equation of state of the form $\omega\equiv
p/\rho$, where $p$ is the spatially homogeneous pressure and
$\rho$ the energy density of the dark energy~\cite{Cai-Wang}. A
value of $\omega<-1/3$ is required for cosmic expansion, and
$\omega=-1$ corresponds to a cosmological
constant~\cite{Carmelli}. The particular case of $\omega=-2/3$ is
extensively analyzed in \cite{Diaz-acc}. A possibility that has
been widely explored, is that of quintessence, a cosmic scalar
field $\phi$ which has not yet reached the minimum of its
potential $V(\phi)$~\cite{Douspis,Wang}. A common example is the
energy of a slowly evolving scalar field with positive potential
energy, similar to the inflaton field used to describe the
inflationary phase of the Universe. In quintessence models the
parameter range is $-1<\omega<-1/3$, and the dark energy decreases
with a scale factor $a(t)$ as $\rho_Q \propto
a^{-3(1+\omega)}$~\cite{Turner,Weinberg}.

However, a note on the choice of the imposition $\omega >-1$ is in
order. This is considered to ensure that the null energy
condition, $T_{\mu\nu}\,k^\mu\,k^\nu>0$, is satisfied, where
$T_{\mu\nu}$ is the stress-energy tensor and $k^{\mu}$ any null
vector. If $\omega<-1$~\cite{Melchiorri,Alcaniz,Hoffman}, a case
certainly not excluded by observation, then the null energy
condition is violated, $\rho+p<0$, and consequently all of the
other energy conditions. Note that the dark energy density is
positive, $\rho>0$. Matter with the property $\omega<-1$ has been
denoted ``phantom energy''. Apart from the null energy condition
violation phantom energy possesses other strange properties,
namely, phantom energy probably mediates a long-range repulsive
force~\cite{Amendola}, phantom thermodynamics leads to a negative
entropy (or negative temparature)~\cite{Odintsov1,Odintsov2}, and
the energy density increases to infinity in a finite
time~\cite{Weinberg,Cald}, at which point the size of the Universe
blows up in a finite time. This is known as the Big Rip. To an
observer on Earth this corresponds to observing the galaxies being
stripped apart, the Earth being itself ripped from its
gravitational attraction to the Sun, before being eventually
ripped apart, followed by the dissociation of molecules and atoms,
and finally of nuclei and nucleons~\cite{Weinberg}. However, it
has been shown that in certain models the presence of phantom
energy does not lead to the above-mentioned
doomsday~\cite{Singh,diaz-phantom,Lopez-Madrid,WuYu,Guo-Zhang,Curbelo}.
It was also shown that quantum corrections may
prevent~\cite{Odintsov3}, or at least delay, the Big Rip by taking
into account the back reaction of conformal quantum fields near
the singularity~\cite{Odintsov4,Odintsov5,Odintsov6}, and the
universe may end up in a deSitter phase before the scale factor
blows up. (It has also been shown that quantum effects, without a
negative kinetic term, could also lead to a super-accelerated
phase of inflation, with a weak energy condition violating, on
average and not just in fluctuations, dark energy equation of
state $\omega<-1$ on cosmological scales~\cite{Onemli1,Onemli2}).
Phantom cosmologies have also been studied in the context of
braneworld scenarios~\cite{Calcagni} and research has been
invested in black holes in an accelerating cosmic expansion with
phantom energy. In an interesting paper it was shown that the
masses of all black holes tend to zero as the phantom energy
universe approaches the Big Rip~\cite{Babichev}. An interesting
candidate for phantom energy, considering a positive energy
density so that the field is purely imaginary, is that of cosmic
axions, described by a rank-3 tensor field predicted by
supergravity and string theories~\cite{diaz-phantom2}. It has also
been shown, recently, that constraints from the current supernovae
Ia Hubble diagram \cite{Majerotto} favor a negative coupling
between dark energy and dark matter, implying the existence of
phantom energy, and an equation of state $w<-1$. In fact, recent
fits to supernovae, cosmic microwave background radiation and weak
gravitational lensing data indicate that an evolving equation of
state $\omega$ which crosses $-1$ is mildly favored
\cite{Zhang1,Ishak}. If confirmed in the future, this behavior has
important implications for theoretical models of dark energy. For
instance, this implies that dark energy is dynamical and, in
addition, to excluding the cosmological constant as a possible
candidate for dark energy, the models with a constant parameter,
such as quintessence and phantom models, cannot de satisfied
either. The evolving dark energy with an equation of state
$\omega$ crossing $-1$ during its evolution was dubbed
``Quintom''~\cite{Zhang1,Zhang2,Zhang3}, as it is different from
the quintessence or phantom fields in the determination of the
evolution and fate of the universe. All of these models present an
extremely fascinating aspect for future experiments focussing on
supernovae, cosmic microwave background radiation and weak
gravitational lensing and for future theoretical research.

As the possibility of phantom energy implies the violation of the
null energy condition, this presents us with a natural scenario
for the existence of traversable wormholes. The latter possess a
peculiar property, namely ``exotic matter'', involving a
stress-energy tensor that violates the null energy
condition~\cite{Morris,Visser}, precisely the fundamental
ingredient of phantom energy. (Recently, wormhole throats were
also analyzed in a higher derivative gravity model governed by the
Einstein-Hilbert Lagrangian, supplemented with $1/R$ and $R^2$
curvature scalar terms~\cite{furey-debened}, and using the
resulting equations of motion, it was found that the weak energy
condition may be respected in the throat vicinity). Thus, one
could imagine an absurdly advanced civilization, for instance,
mining phantom energy necessary to construct and sustain a
traversable wormhole. Another interesting scenario is that due to
the fact of the accelerated expansion of the Universe, macroscopic
wormholes could naturally be grown from the submicroscopic
constructions that originally pervaded the gravitational vacuum,
in a manner similar to the inflating wormholes analyzed by
Roman~\cite{romanLambda}. In fact, Gonz\'{a}lez-D\'{i}az analyzed
the evolution of wormhole and ringhole spacetimes embedded in a
background accelerating Universe~\cite{gonzalez2} driven by dark
energy. It was shown that the wormhole's size increases by a
factor which is proportional to the scale factor of the Universe,
and still increases significantly if the cosmic expansion is
driven by phantom energy. Gonz\'{a}lez-D\'{i}az further considered
the accretion of dark and phantom energy onto Morris-Thorne
wormholes~\cite{diaz-phantom3,diaz-phantom4}. It was shown that
this accretion gradually increases the wormhole throat which
eventually overtakes the accelerated expansion of the universe and
becomes infinite at a time in the future before the big rip. As it
continues accreting phantom energy, the wormhole becomes an
Einstein-Rosen bridge, which pinches off rendering it
non-traversal, and the respective mass decreases rapidly and
vanishes at the big rip.

As the material with the properties necessary to sustain
traversable wormholes, namely, null energy condition violating
phantom energy, probably comprises of 70\% of the constitution of
the Universe, it is of particular interest to investigate the
physical properties and characteristics of these specific wormhole
geometries. In fact, one can trace back to the 1970s examples of
wormholes with a negative kinetic energy
scalar~\cite{homerellis,homerellis2,bronikovWH,kodama}, which may
be considered as the current phantom energy source~\cite{picon}.
In this paper, we will be interested in constructing wormhole
solutions using the equation of state $p=\omega \rho$, with
$\omega <-1$, that describes phantom energy in cosmology.
Independent work was carried out by Sushkov~\cite{Sushkov} (which
we became aware of after the completion of the present paper).
Despite the fact that the notion of phantom energy is that of a
homogeneously distributed fluid in the Universe, as emphasized in
\cite{Sushkov}, it can be extended to inhomogeneous spherically
symmetric spacetimes by regarding that the pressure in the
equation of state $p=\omega \rho$ is now a negative radial
pressure, and noting that the transverse pressure $p_t$ may be
determined from the Einstein field equations. This is
fundamentally the analysis carried out by Sushkov and
Kim~\cite{Sush-Kim}, whilst constructing a time-dependent solution
describing a spherically symmetric wormhole in a cosmological
setting with a ghost scalar field. It was shown that the radial
pressure is negative everywhere and far from the wormhole throat
equals the transverse pressure, showing that the ghost scalar
field behaves essentially as dark energy. Sushkov, in
\cite{Sushkov}, considered specific choices for the distribution
of the energy density threading the wormhole. However, we trace
out a complementary approach by modelling an appropriate wormhole
geometry imposing specific choices for the form function and/or
the redshift function, and consequently determining the
stress-energy components. We find that particularly interesting
solutions exist, and by using the ``volume integral quantifier'',
it is found that these wormhole geometries are, in principle,
sustained by arbitrarily small amounts of averaged null energy
condition (ANEC) violating phantom energy.

This paper is outlined in the following manner. In Section II, we
present a general solution of a traversable wormhole comprising of
phantom energy. We also present the case of a thin shell
surrounding the interior wormhole geometry, in which the surface
stresses obey the phantom energy equation of state. In Section
III, we construct specific traversable wormhole geometries,
namely, asymptotically flat spacetimes and the case of an
isotropic pressure. We also show that using these specific
constructions, and taking into account the ``volume integral
quantifier'', one may theoretically construct these spacetimes
with infinitesimal amounts of ANEC violating phantom energy.
Specific wormhole dimensions and traversal velocities and time are
also determined from the traversability conditions, by considering
a particular wormhole geometry. Finally, in Section IV, we
conclude.

\section{Traversable wormhole spacetimes comprising of phantom energy}

The spacetime metric representing a spherically symmetric and
static wormhole is given by
\begin{equation}
ds^2=-e ^{2\Phi(r)}\,dt^2+\frac{dr^2}{1- b(r)/r}+r^2 \,(d\theta
^2+\sin ^2{\theta} \, d\phi ^2) \label{metricwormhole}\,,
\end{equation}
where $\Phi(r)$ and $b(r)$ are arbitrary functions of the radial
coordinate, $r$. $\Phi(r)$ is denoted as the redshift function,
for it is related to the gravitational redshift; $b(r)$ is called
the form function, because as can be shown by embedding diagrams,
it determines the shape of the wormhole \cite{Morris}. The radial
coordinate has a range that increases from a minimum value at
$r_0$, corresponding to the wormhole throat, to $a$, where the
interior spacetime will be joined to an exterior vacuum solution.

Using the Einstein field equation, $G_{\hat{\mu}\hat{\nu}}=8\pi
\,T_{\hat{\mu}\hat{\nu}}$, in an orthonormal reference frame,
(with $c=G=1$) we obtain the following stress-energy scenario
\begin{eqnarray}
\rho(r)&=&\frac{1}{8\pi} \;\frac{b'}{r^2}   \label{rhoWH}\,,\\
p_r(r)&=&\frac{1}{8\pi} \left[2 \left(1-\frac{b}{r}
\right) \frac{\Phi'}{r} -\frac{b}{r^3}\right]  \label{prWH}\,,\\
p_t(r)&=&\frac{1}{8\pi} \left(1-\frac{b}{r}\right)\left[\Phi ''+
(\Phi')^2- \frac{b'r-b}{2r(r-b)}\Phi'
-\frac{b'r-b}{2r^2(r-b)}+\frac{\Phi'}{r} \right] \label{ptWH}\,,
\end{eqnarray}
in which $\rho(r)$ is the energy density, $p_r(r)$ is the radial
pressure, and $p_t(r)$ is the lateral pressure measured in the
orthogonal direction to the radial direction. Using the
conservation of the stress-energy tensor,
$T^{\hat{\mu}\hat{\nu}}_{\;\;\;\;;\,\hat{\nu}}=0$, we obtain the
following equation
\begin{equation}
p_r'=\frac{2}{r}\,(p_t-p_r)-(\rho +p_r)\,\Phi '
\label{prderivative} \,,
\end{equation}
which can be interpreted as the relativistic Euler equation, or
the hydrostatic equation for equilibrium for the material
threading the wormhole. Note that eq. (\ref{prderivative}) can
also be obtained from the field equations by eliminating the term
$\Phi''$ and taking into account the radial derivative of eq.
(\ref{prWH}).

Now using the equation of state representing phantom energy,
$p_r=\omega \rho$ with $\omega<-1$, and taking into account eqs.
(\ref{rhoWH})-(\ref{prWH}), we have the following condition
\begin{equation}
\Phi'(r)=\frac{b+\omega rb'}{2r^2\,\left(1-b/r \right)} \,.
            \label{EOScondition}
\end{equation}
We now have four equations, namely the field equations, i.e., eqs.
(\ref{rhoWH})-(\ref{ptWH}), and eq. (\ref{EOScondition}), with
five unknown functions of $r$, i.e., $\rho(r)$, $p_r(r)$,
$p_t(r)$, $b(r)$ and $\Phi(r)$. To construct solutions with the
properties and characteristics of wormholes, we consider
restricted choices for $b(r)$ and/or $\Phi(r)$. In particular, by
appropriately choosing the form function, one may integrate eq.
(\ref{EOScondition}), to determine the redshift function,
$\Phi(r)$. As an alternative, one may choose $\Phi(r)$, and
through the following integral
\begin{equation}
b(r)=r_0
\left(\frac{r_0}{r}\right)^{\frac{1}{\omega}}\,e^{-\frac{2}{\omega}\,
\left[\Phi(r)-\Phi(r_0)\right]}\, \left[\frac{2}{\omega}
\int_{r_0}^r\;
\left(\frac{r}{r_0}\right)^{\left(\frac{1+\omega}{\omega}\right)}\,
\Phi'(r)\,e^{\frac{2}{\omega}\,\left[\Phi(r)-\Phi(r_0)\right]}\;dr+1
\right]  \,,
            \label{b:integral}
\end{equation}
obtained from eq. (\ref{EOScondition}), one may determine the form
function, and consequently the stress-energy tensor components.

Now, in cosmology the energy density related to the phantom energy
is considered positive, $\rho>0$, so we shall maintain this
condition. This implies that only form functions of the type
$b'(r)>0$ are allowed. At the throat we have the flaring out
condition given by $(b-b'r)/b^2>0$ \cite{Morris,Visser}, which may
be deduced from the mathematics of embedding. From this we verify
that at the throat $b(r_0)=r=r_0$, the condition $b'(r_0)<1$ is
imposed to have wormhole solutions. For the wormhole to be
traversable, one must demand that there are no horizons present,
which are identified as the surfaces with $e^{2\Phi}\rightarrow
0$, so that $\Phi(r)$ must be finite everywhere. Note that the
condition $1-b/r>0$ is also imposed. We can construct
asymptotically flat spacetimes, in which $b(r)/r\rightarrow 0$ and
$\Phi\rightarrow 0$ as $r\rightarrow \infty$. However, one may
also construct solutions with a cut-off of the stress-energy, by
matching the interior solution to an exterior vacuum spacetime, at
a junction interface. If the junction contains surface stresses,
we have a thin shell, and if no surface stresses are present, the
junction interface is denoted a boundary surface.

For instance, consider that the exterior solution is the
Schwarzschild spacetime, given by
\begin{eqnarray}
ds^2&=&-\left(1-\frac{2M}{r}\right)\,dt^2+
\left(1-\frac{2M}{r}\right)^{-1}dr^2+r^2(d\theta ^2+\sin
^2{\theta}\, d\phi ^2) \label{metricvacuumlambda}  \,.
\end{eqnarray}
In this case the spacetimes given by the metrics eq.
(\ref{metricwormhole}) and (\ref{metricvacuumlambda}) are matched
at $a$, and one has a thin shell surrounding the wormhole. Using
the the Darmois-Israel
formalism~\cite{Sen,Lanczos,Darmois,Israel,Papahamoui}, the
surface stresses are given by
\begin{eqnarray}
\sigma&=&-\frac{1}{4\pi a} \left(\sqrt{1-\frac{2M}{a}}-
\sqrt{1-\frac{b(a)}{a}} \, \right)
    \label{surfenergy}   ,\\
{\cal P}&=&\frac{1}{8\pi a}
\left(\frac{1-\frac{M}{a}}{\sqrt{1-\frac{2M}{a}}}- \zeta \,
\sqrt{1-\frac{b(a)}{a}} \, \right)
    \label{surfpressure}    ,
\end{eqnarray}
where $\zeta=[1+a\Phi'(a)]$ is the redshift parameter
\cite{Lobo-CQG}; $\sigma$ is the surface energy density and ${\cal
P}$ the surface pressure. Construction of wormhole solutions by
matching an interior wormhole spacetime to an exterior vacuum
solution, at a junction surface, were also recently
analyzed~\cite{Lobo-CQG,LLQ,Lobo}. In particular, a thin shell
around a traversable wormhole, with a zero surface energy density
was analyzed in \cite{LLQ}, and with generic surface stresses in
\cite{Lobo-CQG}. A general class of wormhole geometries with a
cosmological constant and junction conditions was explored in
\cite{benedectis}, and a linearized stability analysis of
thin-shell wormholes with $\Lambda$ was studied in \cite{LC-CQG}.
A similar analysis for the plane symmetric case, with a negative
cosmological constant, is done in \cite{LL-PRD}.

The surface mass of the thin shell is given by $M_{\rm s}=4\pi a^2
\sigma$. Note that by rearranging the terms in eq.
(\ref{surfenergy}), one obtains the following relationship
\begin{equation}\label{totalmass}
M=\frac{b(a)}{2}+M_{\rm
s}\left(\sqrt{1-\frac{b(a)}{a}}-\frac{M_{\rm s}}{2a}\right)   \,,
\end{equation}
where $M$ may be interpreted as the total mass of the wormhole in
one asymptotic region.

It is also of interest to impose the phantom energy equation of
state on the surface stresses, i.e., ${\cal P}=\omega \sigma$,
with $\omega<-1$. For this case, we have the following condition
\begin{equation}
(\zeta-2|\omega|)\sqrt{1-\frac{b(a)}{a}}=\frac{\left(2|\omega|
-\frac{1}{2}\right)\,\frac{2M}{a}-(2|\omega|-1)}{\sqrt{1-\frac{M}{a}}}
   \,.
\end{equation}
One has several cases to analyze. If $\zeta \geq 2|\omega|$, then
the thin shell lies in the following range
\begin{equation}
2M<a \leq \left(\frac{2|\omega|
-\frac{1}{2}}{2|\omega|-1}\right)\,2M \,.
\end{equation}
If $\zeta < 2|\omega|$, then $a > 2M(2|\omega|
-1/2)/(2|\omega|-1)$.

If one is tempted to construct a boundary surface, i.e.,
$\sigma={\cal P}=0$, then one needs to impose the conditions:
$b(a)=2M$ and $a = 2M(\zeta -1/2)/(\zeta-1)$. Note that from the
latter condition, taking into account $a>2M$, we have an
additional restriction imposed on the redshift parameter, namely,
$\zeta>1$.

\section{Specific wormhole construction}

\subsection{Asymptotically flat spacetimes}

\subsubsection{$(i)$ Specific choice for the form function:
$b(r)=r_0\left(r/r_0\right)^\alpha$}

Consider the particular choice of the form function
$b(r)=r_0(r/r_0)^\alpha$, with $0<\alpha<1$. For this case we
readily verify that $b'(r)=\alpha(r/r_0)^{\alpha-1}$, so that at
the throat $b'(r_0)=\alpha<1$, and that for $r\rightarrow \infty$
we have $b(r)/r=(r_0/r)^{1-\alpha}\;\rightarrow 0$. From eq.
(\ref{EOScondition}), taking into account the above-mentioned form
function, we find the following solution
\begin{equation}
\Phi(r)=\frac{1}{2}\,\left(\frac{1+\alpha
\omega}{1-\alpha}\right)\,\ln\left[1-\left(\frac{r_0}{r}\right)^{1-\alpha}
\right]  \,.
\end{equation}
The spacetime metric in this case is given by
\begin{equation}
ds^2=-\left[1-(r_0/r)^{1-\alpha} \right]^{\frac{1+\alpha
\omega}{1-\alpha}}\,dt^2+\frac{dr^2}{1- (r_0/r)^{1-\alpha}}+r^2
\,(d\theta ^2+\sin ^2{\theta} \, d\phi ^2) \,.
\end{equation}
Now, if $1+\alpha \omega>0$, then the spacetime is asymptotically
flat, $e^{2\Phi}\rightarrow 1$ as $r\rightarrow \infty$ (recall
that by construction $b/r \rightarrow 0$). However, an event
horizon is located at $r=r_0$, implying the existence of a
non-traversable wormhole.

If $1+\alpha \omega=0$, so that $e^{2\Phi}= 1$, then no event
horizon is present. Thus, for this case the parameters $\alpha$
and $\omega$ are related by the condition $\alpha=-1/\omega$. Note
that this solution can also be obtained in the following manner.
Firstly, consider $\Phi(r)={\rm const}$, and from eq.
(\ref{b:integral}) one obtains $b(r)=r_0(r/r_0)^{-1/\omega}$.
Then, by confronting this solution with the above-mentioned form
function, one may identify the relationship $\alpha=-1/\omega$.
Therefore, if one determines the parameter $\omega$ from
observational cosmology, assuming the existence of phantom energy,
then one may theoretically construct traversable wormholes, by
imposing the condition $\alpha=-1/\omega$ in the above-mentioned
form function and considering a constant redshift function.

Taking into account the notion of the ``volume integral
quantifier'' one may quantify the ``total amount'' of energy
condition violating matter. This notion amounts to calculating the
definite integrals $\int T_{\mu\nu}U^\mu U^\nu \,dV$ and $\int
T_{\mu\nu}k^\mu k^\nu \,dV$, and the amount of violation is
defined as the extent to which these integrals become negative.
(It is also interesting to note that recently, by using the
``volume integral quantifier'', fundamental limitations on ``warp
drive'' spacetimes were found, for non-relativistic warp
velocities~\cite{LV-CQG,LV}). Visser {\it et al} recently found
that by considering specific examples of traversable wormhole
geometries, one may theoretically construct these spacetimes with
infinitesimal amounts of ANEC violating matter~\cite{VKD1,VKD2}.
In this work we will also be interested in investigating whether
this is the case for phantom energy. The integral which provides
information about the ``total amount'' of ANEC violating matter in
the spacetime is given by (see \cite{VKD1,VKD2} for details)
\begin{eqnarray}\label{vol:int}
I_V=\int \left[\rho(r) +p_r(r)\right]\; dV=
\left[(r-b)\,\ln\left(\frac{e^{2\Phi}}{1-b/r}\right)\,\right]^{\infty}_{r_0}-\int_{r_0}^\infty
\left(1-b' \right) \;\left[\ln \left(\frac{e^{2\Phi}}{1-b/r}
\right) \right] \;dr \,.
\end{eqnarray}
Note that the boundary term at the throat $r_0$ vanishes by
construction of the wormhole, and the term at infinity only
vanishes if we are assuming asymptotically flat spacetimes.

Considering the specific choices for the form function and
redshift function for the traversable wormhole, eq.
(\ref{vol:int}) takes the form
\begin{eqnarray}\label{vol:int2}
I_V= \int_{r_0}^\infty \left[1-\alpha \left(\frac{r_0}{r}
\right)^{1-\alpha}\right] \;\left\{\ln \left[1-\left(\frac{r_0}{r}
\right)^{1-\alpha} \right] \right\} \;dr \,.
\end{eqnarray}
Now as in \cite{VKD1,VKD2}, suppose that we have a wormhole field
deviating from the Schwarzschild solution from the throat out to a
radius $a$. In particular, this amounts to matching the interior
solution to an exterior spacetime at $a$. From eqs.
(\ref{surfenergy})-(\ref{surfpressure}), and taking into account
the choices of the form function and the redshift function
considered above, we verify that the junction interface
necessarily comprises of a thin shell. Consider, for simplicity,
$\alpha=1/2$, so that the volume integral assumes the value
\begin{eqnarray}
I_V= r_0 \left(1-\sqrt{\frac{a}{r_0}} \; \right)+a
\left(1-\sqrt{\frac{r_0}{a}} \; \right)\;\left[\ln
\left(1-\sqrt{\frac{r_0}{a}} \; \right) \right]\,.
\end{eqnarray}
Taking the limit $a \rightarrow r_0$, one verifies that $I_V=\int
(\rho + p_r)\; dV \rightarrow 0$. Thus, as in the examples
presented in \cite{VKD1,VKD2}, one may construct a traversable
wormhole with arbitrarily small quantities of ANEC violating
phantom energy. Although this result is not unexpected it is
certainly a fascinating prospect that an advanced civilization may
probably construct and sustain a wormhole with vanishing amounts
of the material that comprises of approximately 70\% of the
constitution of the Universe.

It is also interesting to consider the traversability conditions
required for a human being to journey through the wormhole.
Firstly, the condition required that the acceleration felt by the
traveller should not exceed Earth's gravity, $g_\oplus$, is given
by (see \cite{Morris} for details)
\begin{equation}
\left |\left (1-\frac{b}{r}\right)^{1/2} e^{-\Phi}\,\left(\gamma
\,e^{\Phi}\right)' \right|\leq g_{\oplus} \label{acceleration} \,.
\end{equation}
Secondly, the condition required that the tidal accelerations
should not exceed the Earth's gravitational acceleration yields
the following restrictions
\begin{eqnarray}
\left |\left (1-\frac{b}{r} \right ) \left [\Phi ''+(\Phi ')^2-
\frac{b'r-b}{2r(r-b)}\Phi' \right] \right
|\,\big|\eta^{\hat{1}'}\big| &\leq & g_\oplus   \,,
    \label{radialtidalconstraint}    \\
\left | \frac{\gamma ^2}{2r^2} \left [v^2\left (b'-\frac{b}{r}
\right )+2(r-b)\Phi ' \right] \right | \,\big|\eta^{\hat{2}'}\big|
&\leq &  g_\oplus    \,,    \label{lateraltidalconstraint}
\end{eqnarray}
where $|\eta^{\hat{i}'}|$ is the separation between two arbitrary
parts of the traveller's body, and as in \cite{Morris} we shall
assume $|\eta^{\hat{i}'}|\approx 2\,{\rm m}$ along any spatial
direction in the traveller's reference frame.

Finally, considering that the space stations are positioned just
outside the junction radius, $a$, at $l=-l_1$ and $l=l_2$,
respectively, where $dl=(1-b/r)^{-1/2}\,dr$ is the proper radial
distance, the traversal time as measured by the traveller and for
the observers that remain at rest at space stations are given by
\begin{eqnarray}
\Delta \tau =\int_{-l_1}^{+l_2} \frac{dl}{v\gamma}
\qquad {\rm and} \qquad
\Delta t =\int_{-l_1}^{+l_2} \frac{dl}{v e^{\Phi}}  \,,
\end{eqnarray}
respectively,

We verify that for the wormhole constructed in this Section with
the specific choice of $b(r)=r_0(r/r_0)^\alpha$ and $\Phi(r)={\rm
const}$, considering for simplicity a constant non-relativistic,
$\gamma \approx 1$, traversal velocity, the restriction of the
inequalities (\ref{acceleration}) and
(\ref{radialtidalconstraint}) are readily satisfied. From the
restriction (\ref{lateraltidalconstraint}), evaluated at the
throat one obtains the inequality
\begin{equation}
v \leq
r_0\,\sqrt{\frac{2g_\oplus}{(1-\alpha)\,\big|\eta^{\hat{2}'}\big|}}
\end{equation}
Considering the equality case, taking into account $\alpha=1/2$,
and imposing that the wormhole throat is given by $r_0\approx
10^2\,{\rm m}$, then one obtains $v\approx 4\times 10^2\,{\rm
m/s}$ for the traversal velocity. If one considers that the
junction radius is given by $a\approx 10^4\,{\rm m}$, then from
the traversal times $\Delta \tau \approx \Delta t \approx 2a/v$,
one obtains $\Delta \tau \approx \Delta t \approx 50 \,{\rm s}$.

\subsubsection{$(ii)$ Specific choice for the redshift function:
$\Phi(r)=(r_0/r)$}

Using the redshift function given by $\Phi(r)=(r_0/r)$, and
solving the integral of eq. (\ref{b:integral}), we have as a
solution the form function given by
\begin{equation}\label{form2}
b(r)=2r_0\;\left\{\left(-\frac{2r_0}{\omega
r}\right)^{1/\omega}\;e^{\left(-\frac{2r_0}{\omega
r}\right)}\;\left[C+{\cal
F}\left(\frac{\omega-1}{\omega},-\frac{2r_0}{\omega
r}\right)\right]-1 \right\}   \,,
\end{equation}
where $C$ is a constant, given by
\begin{eqnarray}
C=\frac{3}{2}\left(-\frac{2}{\omega}\right)^{-1/\omega}\,
e^{2/\omega} - {\cal
F}\left(\frac{\omega-1}{\omega},-\frac{2}{\omega}\right)
\end{eqnarray}
and the function ${\cal F}$ is defined as
\begin{eqnarray}
{\cal F}(x,z)&=&\Gamma(x,z)-\Gamma(x) \,,
\end{eqnarray}
where $\Gamma(x)$ and $\Gamma(x,z)$ are the Gamma and the
incomplete Gamma functions, respectively.

Although the form function is extremely messy, the message that
one can extract from this analysis, for $\omega <-1$, is that one
may prove that $b(r)/r \rightarrow 0$ as $r \rightarrow \infty$,
and the condition $1-b/r>0$ is also obeyed. Thus, by construction,
as $\Phi(r) \rightarrow 0$ when $r \rightarrow \infty$, the
spacetime is asymptotically flat.

Using the ``volume integral quantifier'' provided by eq.
(\ref{vol:int}), and substituting the form and redshift functions
chosen in this section, one ends up with an intractable integral.
However, one may prove that for $\omega<-1$, the volume integral
is vanishingly small, i.e., $I_V \rightarrow 0$ as $a \rightarrow
r_0$, and the message is that once again one can theoretically
construct traversable wormholes with vanishing amounts of ANEC
violating phantom energy.

\subsection{Isotropic pressure, $p_r=p_t=p$}

It is of particular interest to consider an isotropic pressure,
$p_r=p_t=p$, so that eq. (\ref{prderivative}) with the equation of
state $p=\omega \rho$, reduces to
\begin{equation}
p'=-\left(\frac{1+\omega}{\omega}\right)p\,\Phi'  \,,
\end{equation}
which is immediately integrated to provide the solution
\begin{equation}
p(r)=\omega \rho(r)=-\frac{1}{8\pi
r_0^2}\,e^{-\left(\frac{1+\omega}{\omega}\right)\,[\Phi(r)-\Phi(r_0)]}
   \label{EOSisotropic}  \,.
\end{equation}
One may consider that $\Phi(r_0)=0$ without a significant loss of
generality. Note that the isotropic pressure is always negative,
implying the presence of an isotropic tension.

Now, substituting eq. (\ref{EOSisotropic}) in eq. (\ref{rhoWH}),
one deduces the following relationship
\begin{equation}
b'(r)=-\frac{1}{\omega}\left(\frac{r}{r_0}\right)^2\;
e^{-\left(\frac{1+\omega}{\omega}\right)\,\Phi(r)}
         \,,   \label{b:derivative}
\end{equation}
which may be rewritten as
\begin{equation}
\Phi(r)=-\left(\frac{\omega}{1+\omega}\right)\ln\left[-\omega
b'(r) \left(\frac{r_0}{r}\right)^2 \right]
      \,.   \label{redshift}
\end{equation}
From this relationship one verifies that for $\Phi(r)$ to be
finite, then $b(r)\propto r^3$, so that generically one cannot
construct asymptotically flat traversable wormholes with isotropic
pressures. Nevertheless, one may match the interior wormhole
solution to an exterior vacuum spacetime at a finite junction
surface.


One may consider specific choices for the redshift function, then
from eq. (\ref{b:derivative}), deduce $b(r)$. However, using the
form function considered above, i.e., $b(r)=r_0\,(r/r_0)^\alpha$,
with $0<\alpha<1$, one finds that the redshift function, eq.
(\ref{redshift}), is given by
\begin{equation}
\Phi(r)=\ln\left[\left(-\frac{1}{\omega\,\alpha}\right)^
{\left(\frac{\omega}{1+\omega}\right)}\left(\frac{r}{r_0}\right)^{\omega
\,\left(\frac{3-\alpha}{1+\omega}\right)} \right] \,.
\end{equation}
As we have considered that $\Phi(r_0)=0$, then the relationship
$\alpha=-1/\omega$ is imposed. Thus, the spacetime metric is given
by
\begin{equation}
ds^2=-(r/r_0)^{2\omega
\,\left(\frac{3-\alpha}{1+\omega}\right)}\,dt^2+\frac{dr^2}{1-
(r_0/r)^{1-\alpha}}+r^2 \,(d\theta ^2+\sin ^2{\theta} \, d\phi ^2)
\,,
\end{equation}
which, as generically noted above, is not asymptotically flat, so
we need to match this solution to an exterior vacuum spacetime.
The stress-energy scenario is given by
\begin{equation}
p(r)=\omega \rho=-\frac{1}{8\pi
r_0^2}\;\left(\frac{r_0}{r}\right)^{3-\alpha}  \,.
\end{equation}

Using the ``volume integral quantifier'', eq. (\ref{vol:int}),
with a cut-off of the stress-energy at $a$, and taking into
account the choices for the form function and redshift function
considered above, we have
\begin{eqnarray}\label{vol:int3}
I_V&=&
a\left[1-\frac{b(a)}{a}\right]\,\left\{\ln\left[\frac{e^{2\Phi(a)}}{1-b(a)/a}\right]\right\}-\int_{r_0}^a
\left(1-b' \right) \;\left[\ln \left(\frac{e^{2\Phi}}{1-b/r}
\right) \right] \;dr
        \\
&=& a\left[1-\left(\frac{r_0}{a} \right)^{1-\alpha}\right]
\;\left\{\ln \left[\frac{
\left(a/r_0
\right)^{2\omega\left(\frac{3-\alpha}{1+\omega}\right)}}{1-\left(r_0/a
\right)^{1-\alpha}} \right] \right\}
           -\int_{r_0}^a \left[1-\alpha \left(\frac{r_0}{r}
\right)^{1-\alpha}\right] \;\left\{\ln \left[\frac{\left(r/r_0
\right)^{2\omega\left(\frac{3-\alpha}{1+\omega}\right)}}{1-\left(r_0/r
\right)^{1-\alpha}} \right] \right\} \;dr \,.
\end{eqnarray}
Considering the specific example of $\omega=-1/\alpha=-2$, the
volume integral takes the following value
\begin{eqnarray}
I_V= a \left(1-\sqrt{r_0/a} \; \right)\;\ln
\left[\frac{(a/r_0)^{10}}{1-\sqrt{r_0/a}} \; \right]
+\left(10a+11r_0-21r_0\sqrt{a/r_0} \; \right)+a
\left(\sqrt{r_0/a}-1 \; \right)\;\ln
\left[\frac{(a/r_0)^{21/2}}{\sqrt{a/r_0}-1} \; \right] \,.
\end{eqnarray}
Once again taking the limit $a \rightarrow r_0$, one verifies that
$I_V \rightarrow 0$, and as before one may construct a traversable
wormhole with arbitrarily small quantities of ANEC violating
phantom energy.

\section{Summary and Discussion}

A possible candidate for the accelerated expansion of the Universe
is ``phantom energy'', a cosmic fluid governed by an equation of
state of the form $\omega=p/\rho<-1$, which consequently violates
the null energy condition. This property is precisely the
fundamental ingredient needed to sustain traversable wormholes. It
was shown in \cite{diaz-phantom3} that as a (submicroscopic)
wormhole accretes phantom energy, the radius of the respective
throat will gradually increase to macroscopic dimensions. Thus, it
seems that as the phantom energy dominates, a natural process for
the formation and growth of macroscopic traversable wormholes
exists. As the Universe probably constitutes of approximately 70\%
of phantom energy, one may also imagine an absurdly advanced
civilization mining this cosmic fluid to construct and maintain
traversable wormhole geometries with the characteristics described
in this paper. We have analyzed the physical properties and
characteristics of traversable wormholes using the specific
phantom energy equation of state. We have constructed a thin shell
around the interior wormhole solution, by imposing the phantom
energy equation of state on the surface stresses. We have also
analyzed specific wormhole geometries, by considering
asymptotically flat spacetimes and by imposing an isotropic
pressure. Using the ``volume integral quantifier'' we have
verified that it is theoretically possible to construct these
geometries with vanishing amounts of ANEC violating phantom
energy. Specific wormhole dimensions and the traversal velocity
and time were also deduced from the traversability conditions for
a particular wormhole geometry.

These phantom energy traversable wormholes have far-reaching
physical and cosmological implications. Apart from being used for
interstellar shortcuts, an absurdly advanced civilization may
convert them into time-machines~\cite{mty,Visser,Kluwer}. This is
a troubling issue, depending on one's point of view, as it
probably implies the violation of causality. The cosmological
implications are also extremely interesting. As the wormhole
continues to accrete phantom energy, the throat will blow up in a
finite time $\tilde{t}$ before the occurrence of the big rip
singularity at $t_*$~\cite{diaz-phantom3}. At $\tilde{t}$, the
exotic energy density becomes zero, and the traversable wormhole
is converted into a non-traversable Einstein-Rosen bridge, which
pinches off producing a black hole/white hole pair, in which the
mass tends to zero as the big rip singularity is approached
\cite{diaz-phantom3}. In \cite{gonzalez2} it was shown that the
solution of the scale factor derived from the Friedamnn equation
shows two branches around the occurrence of the big rip, $t_*$.
From the first, one verifies that the universe accelerates towards
the big rip singularity at $t_*$, and the other solution, $t>t_*$,
describes a universe which exponentially decelerates towards a
zero size as $t \to \infty$. Therefore, in a rather speculative
scenario, one may imagine a grown macroscopic wormhole with one
mouth opening in the expanding universe and the other in the
contracting universe. As the first mouth is expanding and the
second contracting, a time-shift would be created between both
mouths, transforming the wormhole into a time machine, so that a
traveller journeying through the wormhole before the big rip would
be transported into his future, and thus circumvent the big rip
singularity. Finally, we point out that the confirmation of the
existence of phantom energy, with an equation of state $w<-1$,
from observational cosmology is an extremely fascinating aspect
for future experiments focussing on supernovae, cosmic microwave
background radiation and weak gravitational lensing, and
consequently for theoretical research.

\section*{Acknowledgements}

We thank Sayan Kar and Cristian Armend\'ariz-Pic\'on for pointing
out the association of the H. Ellis drainhole with phantom energy;
S. Odintsov for pointing out that quantum corrections may prevent,
or at least delay, the Big Rip; Xinmin Zhang for calling to our
attention the models with an equation of state crossing $-1$; and
Pedro Gonz\'{a}lez-D\'{i}az for helpful comments.


\begin{thebibliography}{99}


\bibitem{Riess2}
A. Grant {\it et al}, ``The Farthest known supernova: Support for
an accelerating Universe and a glimpse of the epoch of
deceleration,'' Astrophys. J. {\bf 560} 49-71 (2001)
[arXiv:astro-ph/0104455].

\bibitem{Perlmutter}
S. Perlmutter, M. S. Turner and M. White, ``Constraining dark
energy with SNe Ia and large-scale strucutre,'' Phys. Rev. Lett.
{\bf 83} 670-673 (1999) [arXiv:astro-ph/9901052].

\bibitem{Bennet}
C. L. Bennett {\it et al}, ``First year {\it Wilkinson Microwave
Anisotropy Probe} (WMAP) observations: Preliminary maps and basic
results,'' Astrophys. J. Suppl. {\bf 148} 1 (2003)
[arXiv:astro-ph/0302207].

\bibitem{Hinshaw}
G. Hinshaw {\it et al}, ``First year {\it Wilkinson Microwave
Anisotropy Probe} (WMAP) observations: The angular power
spectrum,'' [arXiv:astro-ph/0302217].


\bibitem{Cai-Wang}
R. Cai and A. Wang , ``Cosmology with Interaction between Phantom
Dark Energy and Dark Matter and the Coincidence Problem,''
[arXiv:hep-th/0411025].


\bibitem{Carmelli}
M. Carmelli, ``Accelerating universe, cosmological constant and
dark energy,'' [arXiv:astro-ph/0111259].


\bibitem{Diaz-acc}
P. F. Gonz\'{a}lez-D\'{i}az, ``Stable accelerating universe with
no hair,'' Phys. Rev. D {\bf 65} 104035 (2002)
[arXiv:hep-th/0203210].



\bibitem{Douspis}
M. Douspis, A. Riazuelo, Y. Zolnierowski and A. Blanchard,
``Cosmological parameters estimation in the Quintessence
Paradigm,'' Astrophys. J. Suppl. {\bf 148} 135 (2003)
[arXiv:astro-ph/0212097].

\bibitem{Wang}
L. Wang, R. R. Caldwell, J. P. Ostriker and P. J. Steinhardt,
``Cosmic Concordance and Quintessence,'' Astrophys. J. {\bf 530}
17-35 (2000) [arXiv:astro-ph/9901388].


\bibitem{Turner}
M. S. Turner, ``Dark energy and the New Cosmology,''
[arXiv:astro-ph/0108103].


\bibitem{Weinberg}
R. R. Caldwell, M. Kamionkowski and N. N. Weinberg, ``Phantom
Energy and Cosmic Doomsday,'' Phys. Rev. Lett. {\bf 91} 071301
(2003) [arXiv:astro-ph/0302506].

\bibitem{Melchiorri}
A. Melchiorri, L. Mersini, C. J. \"{O}dman and M. Trodden, ``The
state of the dark energy equation of state,'' Phys. Rev. D {\bf
68} 043509 (2003) [arXiv:astro-ph/0211522].

\bibitem{Alcaniz}
J. S. Alcaniz, ``Testing dark energy beyond the cosmological
constant barrier,'' Phys. Rev. D {\bf 69} 083521 (2004)
[arXiv:astro-ph/0312424].

\bibitem{Hoffman}
S. M. Carroll, M. Hoffman and M. Trodden, ``Can the dark energy
equation-of-state parameter $w$ be less than $-1$?,'' Phys. Rev. D
{\bf 68} 023509 (2003) [arXiv:astro-ph/0301273].


\bibitem{Amendola}
L. Amendola, ``Phantom energy mediates a long-range repulsive
force,'' Phys. Rev. Lett. {\bf 93} 181102 (2004)
[arXiv:hep-th/0409224].

\bibitem{Odintsov1}
I. Brevik, S. Nojiri, S. D. Odintsov and L. Vanzo, ``Entropy and
universality of Cardy-Verlinde formula in dark energy universe,''
Phys. Rev. D {\bf 70} 043520 (2004) [arXiv:hep-th/0401073].

\bibitem{Odintsov2}
S. Nojiri and S. D.Odintsov, ``The final state and thermodynamics
of dark energy universe,'' Phys. Rev. D {\bf 70} 103522 (2004)
[arXiv:hep-th/0408170].




\bibitem{Cald}
R. R. Caldwell, ``A phantom menace: Cosmological consequences of a
dark energy component with super-negative equation of state,''
Phys. Lett. {\bf B545} 23-29 (2002) [arXiv:astro-ph/9908168].


\bibitem{Singh}
P. Singh, M. Sami and N. Dadhich, ``Cosmological Dynamics of
Phantom Field,'' Phys.Rev. D {\bf 68} 023522 (2003)
[arXiv:hep-th/0305110].

\bibitem{diaz-phantom}
P. F. Gonz\'{a}lez-D\'{i}az, ``You need not be afraid of phantom
energy,'' Phys. Rev. D {\bf 68} 021303(R) (2003)
[arXiv:astro-ph/0305559].


\bibitem{Lopez-Madrid}
M. Bouhmadi-Lopez and J. A. Jimenez Madrid, ``Escaping the Big
Rip?,'' [arXiv:astro-ph/0404540].

\bibitem{WuYu}
P. Wu and H. Yu, ``Avoidance of Big Rip In Phantom Cosmology by
Gravitational Back Reaction,'' [arXiv:astro-ph/0407424].

\bibitem{Guo-Zhang}
Z. Guo and Y. Zhang, ``Interacting Phantom Energy,'' Phys. Rev. D
{\bf 71} 023501 (2005) [arXiv:astro-ph/0411524].

\bibitem{Curbelo}
R. Curbelo, T. Gonzalez and I. Quiros, ``Interacting Phantom
Energy and Avoidance of the Big Rip Singularity,''
[arXiv:astro-ph/0502141].

\bibitem{Odintsov3}
E. Elizalde, S. Nojiri and S. D.Odintsov, ``Late-time cosmology in
(phantom) scalar-tensor theory: dark energy and the cosmic
speed-up,'' Phys. Rev. D {\bf 70} 043539 (2004)
[arXiv:hep-th/0405034].

\bibitem{Odintsov4}
S. Nojiri and S. D.Odintsov, ``Quantum escape of sudden future
singularity,'' Phys. Lett. {\bf B595} 1-8 (2004)
[arXiv:hep-th/0405078].

\bibitem{Odintsov5}
S. Nojiri, S. D.Odintsov and  S. Tsujikawa, ``Properties of
singularities in (phantom) dark energy universe,''
[arXiv:hep-th/0501025].

\bibitem{Odintsov6}
E. Elizalde, S. Nojiri, S. D.Odintsov and  P. Wang, ``Dark Energy:
Vacuum Fluctuations, the Effective Phantom Phase, and
Holography,'' [arXiv:hep-th/0502082].


\bibitem{Onemli1}
V. K. Onemli and R. P. Woodard, ``Super-acceleration from
massless, minimally coupled $\phi^4$,'' Class. Quant. Grav. {\bf
19} 4607 (2002) [arXiv:gr-qc/0204065].

\bibitem{Onemli2}
V. K. Onemli and R. P. Woodard, ``Quantum effects can render
$\omega<-1$ on cosmological scales,'' Phys. Rev. D {\bf 70} 107301
(2004) [arXiv:gr-qc/0406098].

\bibitem{Calcagni}
G. Calcagni, ``Patch dualities and remarks on nonstandard
cosmologies,'' Phys. Rev. D {\bf 71}, 023511 (2005)
[arXiv:gr-qc/0410027].

\bibitem{Babichev}
E. Babichev, V. Dokuchaev and Yu. Eroshenko , ``Black hole mass
decreasing due to phantom energy accretion,'' Phys. Rev. Lett.
{\bf 93} 021102 (2004) [arXiv:gr-qc/0402089].

\bibitem{diaz-phantom2}
P. F. Gonz\'{a}lez-D\'{i}az, ``Axion phantom energy,'' Phys. Rev.
D {\bf 69} 063522 (2004) [arXiv:hep-th/0401082].

\bibitem{Majerotto}
E. Majerotto, D. Sapone and L. Amendola, ``Supernovae type Ia data
favour negatively coupled phantom energy,''
[arXiv:astro-ph/0410543].

\bibitem{Zhang1}
B. Feng, X. Wang and X. Zhang, ``Dark energy constraints from the
cosmic age and supernovae,'' Phys. Lett. {\bf B607} 35-41 (2005)
[arXiv:astro-ph/0404224].

\bibitem{Ishak}
A. Upadhye, M. Ishak and P. J. Steinhardt, ``Dynamical dark
energy: Current constraints and forecasts,''
[arXiv:astro-ph/0411803].

\bibitem{Zhang2}
Z. Guo, Y. Piao, X. Zhang and Y. Zhang, ``Cosmological evolution
of a quintom model of dark energy,'' Phys. Lett. {\bf B608}
177-182 (2005) [arXiv:astro-ph/0410654].

\bibitem{Zhang3}
X.F. Zhang, H. Li, Y. Piao Z. and X. Zhang and X. Zhang,
``Two-field models of dark energy with equation of state across
$-1$,'' [arXiv:astro-ph/0501652].


\bibitem{Morris}
M. S. Morris and K. S. Thorne, ``Wormholes in spacetime and their
use for interstellar travel: A tool for teaching General
Relativity,'' Am. J. Phys. {\bf 56}, 395 (1988).

\bibitem{Visser}
Visser M 1995 {\it Lorentzian Wormholes: From Einstein to Hawking}
(American Institute of Physics, New York)

\bibitem{furey-debened}
N. Furey and A. DeBenedictis, ``Wormhole throats in $R^m$
gravity,'' Class.\ Quant.\ Grav.\ {\bf 22}, 313 (2005)
[arXiv:gr-qc/0410088].


\bibitem{romanLambda} T. A. Roman, ``Inflating Lorentzian wormholes,''
Phys. Rev. D {\bf 47}, 1370 (1993) [arXiv:gr-qc/9211012].



\bibitem{gonzalez2}
P. F. Gonz\'alez-D\'{\i}az, ``Wormholes and ringholes in a
dark-energy universe,'' Phys. Rev. D {\bf 68}, 084016 (2003)
[arXiv:astro-ph/0308382].

\bibitem{diaz-phantom3}
P. F. Gonz\'{a}lez-D\'{i}az, ``Achronal cosmic future,'' Phys.
Rev. Lett. {\bf 93} 071301 (2004) [arXiv:astro-ph/0404045].

\bibitem{diaz-phantom4}
P. F. Gonz\'{a}lez-D\'{i}az and J. A. Jimenez-Madrid, ``Phantom
inflation and the `Big Trip','' Phys. Lett. {\bf B596} 16-25
(2004) [arXiv:hep-th/0406261].

\bibitem{homerellis} H. G. Ellis, ``Ether flow through a drainhole:
A particle model in general relativity,'' J. Math. Phys. {\bf 14},
104 (1973).

\bibitem{homerellis2} H. G. Ellis, ``The evolving, flowless drain hole:
a nongravitating particle model in general relativity theory,''
Gen. Rel. Grav. {\bf 10}, 105-123 (1979).

\bibitem{bronikovWH} K. A. Bronnikov, ``Scalar-tensor theory
and scalar charge,'' Acta Phys. Pol. B {\bf 4}, 251 (1973).

\bibitem{kodama} T. Kodama, ``General-relativistic nonlinear field:
A kink solution in a generalized geometry,'' Phys. Rev. D {\bf
18}, 3529 (1978).

\bibitem{picon}
C. Armend\'ariz-Pic\'on, ``On a class of stable, traversable
Lorentzian wormholes in classical general relativity''. Phys. Rev.
D {\bf 65}, 104010 (2002) [arXiv:gr-qc/0201027].

\bibitem{Sushkov}
S.~Sushkov, ``Wormholes supported by a phantom energy,'' Phys.
Rev. D {\bf 71}, 043520 (2005) [arXiv:gr-qc/0502084].

\bibitem{Sush-Kim}
S. V. Sushkov and S. W. Kim, ``Cosmological evolution of a ghost
scalar field,'' Gen. Rel. Grav. {\bf 36}, 1671 (2004)
[arXiv:gr-qc/0404037].

\bibitem{Sen}
N. Sen, ``\"{U}ber die grenzbedingungen des schwerefeldes an
unsteig keitsfl\"{a}chen,'' Ann. Phys. (Leipzig) {\bf 73}, 365
(1924).

\bibitem{Lanczos}
K. Lanczos, ``Fl\"{a}chenhafte verteiliung der materie in der
Einsteinschen gravitationstheorie,'' Ann. Phys. (Leipzig) {\bf
74}, 518 (1924).

\bibitem{Darmois}
G. Darmois, ``M\'emorial des sciences math\'ematiques XXV,''
Fasticule XXV ch V (Gauthier-Villars, Paris, France, 1927).

\bibitem{Israel}
W. Israel, ``Singular hypersurfaces and thin shells in general
relativity,''   Nuovo Cimento {\bf 44}B, 1 (1966); and corrections
in {\it ibid.} {\bf 48}B, 463 (1966).

\bibitem{Papahamoui}
A. Papapetrou and A. Hamoui, ``Couches simple de mati\`ere en
relativit\'e g\'en\'erale,''  Ann. Inst. Henri Poincar\'e {\bf 9},
179 (1968).


\bibitem{Lobo-CQG}
F. S. N. Lobo, ``Surface stresses on a thin shell surrounding a
traversable wormhole,'' Class. Quant. Grav. {\bf 21} 4811 (2004)
[arXiv:gr-qc/0409018].

\bibitem{LLQ}
J. P. S. Lemos, F. S. N. Lobo and S. Q. de Oliveira,
``Morris-Thorne wormholes with a cosmological constant,'' Phys.
Rev. D {\bf 68}, 064004 (2003) [arXiv:gr-qc/0302049].


\bibitem{Lobo}
F. S. N. Lobo, ``Energy conditions, traversable wormholes and dust
shells,'' [arXiv:gr-qc/0410087].

\bibitem{benedectis}
A. DeBenedictis and A. Das, ``On a general class of wormholes,''
Class. Quant. Grav. {\bf 18}, 1187 (2001) [arXiv:gr-qc/0009072].

\bibitem{LC-CQG}
F. S. N. Lobo and P. Crawford, ``Linearized stability analysis of
thin-shell wormholes with a cosmological constant,'' Class. Quant.
Grav. {\bf 21}, 391 (2004) [arXiv:gr-qc/0311002].

\bibitem{LL-PRD}
J.~P.~S.~Lemos and F.~S.~N.~Lobo, ``Plane symmetric traversable
wormholes in an anti-de Sitter background,'' Phys.\ Rev.\ D {\bf
69} (2004) 104007 [arXiv:gr-qc/0402099].


\bibitem{LV-CQG}
F.~S.~N.~Lobo and M.~Visser, ``Fundamental limitations on `warp
drive' spacetimes,'' Class.\ Quant.\ Grav.\ {\bf 21}, 5871 (2004).
[arXiv:gr-qc/0406083].

\bibitem{LV}
F.~S.~N.~Lobo and M.~Visser, ``Linearized warp drive and the
energy conditions,'' [arXiv: gr-qc/0412065].


\bibitem{VKD1}
M. Visser, S. Kar and N. Dadhich, ``Traversable wormholes with
arbitrarily small energy condition violations,'' Phys. Rev. Lett.
{\bf 90}, 201102 (2003) [arXiv:gr-qc/0301003].


\bibitem{VKD2}
S.~Kar, N.~Dadhich and M.~Visser, ``Quantifying energy condition
violations in traversable wormholes,'' Pramana {\bf 63}, 859-864
(2004) [arXiv:gr-qc/0405103].


\bibitem{mty}
M. S. Morris, K. S. Thorne and U. Yurtsever, ``Wormholes, Time
Machines and the Weak Energy Condition,'' Phy. Rev. Lett. {\bf
61}, 1446 (1988).

\bibitem{Kluwer}
For a review article, see for example:

F.~Lobo and P.~Crawford, ``Time, Closed Timelike Curves and
Causality,'' in The Nature of Time: Geometry, Physics and
Perception, NATO Science Series II. Mathematics, Physics and
Chemistry - Vol. {\bf 95}, Kluwer Academic Publishers, R. Buccheri
et al. eds, pp.289-296 (2003) [arXiv:gr-qc/0206078].



\end{thebibliography}
\end{document}